\documentclass[aps,prl,twocolumn,superscriptaddress,floatfix,bibnotes,amsmath,amssymb,nolongbibliography]{revtex4-2}
\usepackage{adjustbox}
\usepackage{graphicx}% Include figure files
\usepackage{dcolumn}% Align table columns on decimal point
\usepackage{bm}% bold math
\usepackage{xcolor}
\usepackage[inkscapearea=page]{svg}
\DeclareMathOperator{\Tr}{Tr}
\usepackage{soul}
\usepackage[T1]{fontenc}
\usepackage{amstext}
\usepackage{amsmath}
\usepackage{amsfonts}
\usepackage{amssymb}
\usepackage{float}
\usepackage{algorithm}
\usepackage{algpseudocode}
\usepackage{appendix}
\usepackage{multirow}
\usepackage{physics}
\usepackage{hyperref}

\begin{document}
\preprint{APS/123-QED}

\title{Enhancing Quantum Circuit Noise Robustness from a Geometric Perspective}% Force line breaks with \\
%\thanks{A footnote to the article title}%
\author{Junkai Zeng}
\email{zengjk@sustech.edu.cn}
\affiliation{Shenzhen Institute for Quantum Science and Engineering (SIQSE), Southern University of Science and Technology, Shenzhen, P. R. China}%
\affiliation{International Quantum Academy (SIQA), and Shenzhen Branch, Hefei National Laboratory, Futian District, Shenzhen, P. R. China}
\author{Yong-Ju Hai}
\affiliation{International Quantum Academy (SIQA), and Shenzhen Branch, Hefei National Laboratory, Futian District, Shenzhen, P. R. China}
\author{Hao Liang}
\affiliation{Shenzhen Institute for Quantum Science and Engineering (SIQSE), Southern University of Science and Technology, Shenzhen, P. R. China}%
\affiliation{International Quantum Academy (SIQA), and Shenzhen Branch, Hefei National Laboratory, Futian District, Shenzhen, P. R. China}
\author{Xiu-Hao Deng}
\email{dengxh@sustech.edu.cn}
\affiliation{Shenzhen Institute for Quantum Science and Engineering (SIQSE), Southern University of Science and Technology, Shenzhen, P. R. China}%
\affiliation{International Quantum Academy (SIQA), and Shenzhen Branch, Hefei National Laboratory, Futian District, Shenzhen, P. R. China}
\date{\today}% It is always \today, today,
             %  but any date may be explicitly specified

\begin{abstract}

Quantum errors in noisy environments remain a major obstacle to advancing quantum information technology. 
In this work, we expand a recently developed geometric framework, originally utilized for analyzing noise accumulation and creating dynamical error-correcting gates at the control pulse level, to now study noise dynamics at the quantum circuit level.
 Through a geometric perspective, we demonstrate how circuit noise robustness can be enhanced using twirling techniques. 
 Additionally, we show that circuits modified by random twirling correspond to random walk trajectories in this geometric framework, and provide a fresh perspective on randomized compiling by analytically deriving the perturbative expression for the resultant Pauli noise channel.
 We also illustrate that combining robustness optimization strategies at both the control pulse and circuit levels can significantly boost overall circuit fidelity even further through numerical examples.
This research illuminates pathways to achieving noise-resistant quantum control beyond mere optimization of control pulses.

% Quantum errors resulting from noisy environments pose a significant challenge to the advancement of quantum information technology. To address this issue, we propose a geometric approach that unifies noise dynamics treatments at both pulse and circuit levels for error correction or mitigation. Our method enables robust circuit compilation to deterministically tailor noises to enhance the fidelity of GHZ states. Using geometric diagrams, we illustrate how coherent noises can be transformed into random walks to effectively mitigate errors. We also showcase the incorporation of quantum gates with robust pulses into circuits to further improve fidelity and robustness. Our results demonstrate the potential of our geometric approach to advance quantum information processing.
\end{abstract}

\maketitle

\textit{Introduction.}- Quantum technology has the potential to outperform classical systems in certain computing tasks\cite{arute2019quantum,wu2021strong,daley2022practical}. Still, it faces significant challenges due to noise from various sources, which can jeopardize the outcomes of quantum circuits. This has motivated the development of robust control techniques to discover robust pulse waveforms that ensure high-fidelity quantum operations in noisy environments~\cite{bentleyNumericOptimizationConfigurable2020,codyjonesDynamicalDecouplingQubit2012,edmundsDynamicallyCorrectedGates2020,greenArbitraryQuantumControl2013,hansenPulseEngineeringGlobal2021,kabytayevRobustnessCompositePulses2014,khodjastehDesigningPracticalHighfidelity2013,khodjastehArbitrarilyAccurateDynamical2010,khodjastehDynamicallyErrorCorrectedGates2009,khodjastehFaultTolerantQuantumDynamical2005,sauvageOptimalControlFamilies2022}. 
These techniques have led to substantial gate fidelity improvements in various platforms such as semiconductor spin qubits\cite{veldhorst2015two,huang2019fidelity,xue2022quantum,mills2022two}, superconducting qubits\cite{sheldon2016procedure,krantz2019quantum,werninghaus2021leakage}, and trapped ions\cite{egan2021fault,pogorelov2021compact}. Nevertheless, it is crucial to acknowledge that some circuits inherently resist noise, in a manner that is independent of gate realization at the pulse level. For instance, dynamical decoupling gate sequences demonstrate robustness against coherent errors despite being composed of imperfect gates. By embedding noise resistance at the circuit level, it's possible to achieve robustness with potentially simpler pulse designs while still maintaining high circuit fidelity. Despite its potential, this approach to noise robustness at the circuit level remains largely unexplored. 

To enhance the robustness at the pulse level, a geometric framework known as space-curve quantum control has been developed for crafting robust general qubit gates~\cite{zeng2018general,zeng2018fastest,zhuangNoiseresistantLandauZenerSweeps2022,dongDoublyGeometricQuantum2021,buterakosGeometricalFormalismDynamically2021,barnesDynamicallyCorrectedGates2022, hai2022universal}. By mapping noisy qubit dynamics to geometric trajectories in the space spanned by operator basis, one can study how noises impact qubits and how control Hamiltonian fixes the errors. These trajectories together as the Quantum Erroneous Evolution Diagram (QEED) \cite{hai2022universal} illustrate the accumulation dynamics of quantum errors via different noise channels driven by different control Hamiltonians. Based on this framework, the manuscript expands the analysis from individual pulses to entire quantum circuits to assess how different gate sequences affect error propagation. The same quantum circuit can be configured with different gate sequences, each linked to a unique QEED. This understanding offers a method to manage noise dynamics and improve circuit accuracy through optimized compilations of gate sequences. Earlier efforts toward the same goal include the dynamically error-corrected gate with sequences of quantum gates~\cite{khodjastehDynamicallyErrorCorrectedGates2009}, and randomized compiling (RC) to mitigate the coherent errors in quantum circuits~\cite{wallmanNoiseTailoringScalable2016,hashimRandomizedCompilingScalable2021,guNoiseresilientPhaseEstimation2022,urbanek2021mitigating}.

A key strategy for reconfiguring quantum circuits into their equivalence is inserting twirling gates and their complements, which alters the error accumulation and propagation. Employing QEED as a visual tool, we advocate for constructing robust quantum circuits that preemptively correct errors through carefully designed twirling gate sequences. Moreover, we demonstrate that QEEDs associated with circuits modified through randomized twirling sequences correspond to random walk trajectories, and averaging across multiple random circuit realizations through RC can mitigate errors effectively. Our findings further show that incorporating robust optimal control methodologies at the pulse level with circuit-level error mitigation significantly enhances quantum circuit fidelity, as evidenced by our numerical analyses. Our formalism provides a deeper understanding of the propagation of quantum error and opens up opportunities for developing more noise-resistant circuit compiling protocols beyond RC.

\textit{Geometric Formalism}.- We introduce geometric formalism by first studying generic noisy quantum dynamics and then extending our analysis to circuit models. Consider a multi-qubit system subject to coherent noise, described by Hamiltonian: $H(t)=H_{0}(t)+\sum_{i}\epsilon _{i}(t)V_{i}$. Here $H_{0}$ is the noise-free part of the Hamiltonian, $V_{i}$ the noise operator, and $\epsilon _{i}(t)$ the noise strength. The time dependency of the noise $\epsilon _{i}(t)$ can take various forms, such as constant, randomly fluctuating, drifting, or proportional to control amplitude, depending on the noise source. For coherent noise, $V_{i}$ is typically the tensor product of Pauli matrices for qubit systems.

The evolution operator can be decomposed into two parts: 
\begin{equation}
U(t)=U_{0}(t)U_{\epsilon }(t),  \label{eq:decom}
\end{equation}%
where the error-free evolution $U_{0}(t)=\mathcal{T}e^{-i\int dsH_{o}(s)}$ and the error evolution $U_{\epsilon }(t)=\mathcal{T}e^{-i\Phi (t)}$, with the $\Phi (t)=\int_{0}^{t}ds\sum_{i}\epsilon _{i}(s)U_{0}^{\dagger}(s)V_{i}U_{0}(s)$ referred to as \textit{error operator} in the following. As the noise $\epsilon _{i}(t)$ is generally weak, we apply Magnus expansion and truncate higher order terms so that $U_{\epsilon }(t)\approx e^{-i\Phi (t)}$. Since general $n$-qubit operators can be expanded using Pauli basis $\sigma_{j}\in span\{I,X,Y,Z\}^{\otimes n}$, the error operator can be written as
\begin{equation}
\Phi (t)=\sum_{ij}\sigma_{j}\int_{0}^{t}\epsilon
_{i}(s)T_{j}^{(i)}(s)ds\equiv \sum_{ij}\sigma_{j}\mathbf{R}%
_{j}^{(i)}(t)  \label{eq:timedependent}
\end{equation}%
%Integration by parts is used to derive the second equation.
Here $T_{j}^{(i)}(t)=\frac{1}{D}\Tr\left( \sigma_{j}\int_{0}^{t}U_{0}^{\dagger }(s)V_{i}U_{0}(s)ds\right) $ represents the first-order susceptibility of operator $\sigma_{j}$ to noise component $V_{i}$, and $D$ is the rank of $V_i$. The vector function $\mathbf{R^{(i)}}(t)$ characterizes how much error accumulated on each operator basis $\sigma_{j}$ at time $t$. Geometrically, a point at $\mathbf{R^{(i)}}(t)$ moves in a high dimensional Euclidean space with speed $\epsilon _{i}(t)$ and direction $\mathbf{T^{(i)}}(t)$, for normalized $V_{i}$. Hence, the trajectory of this point forms a geometric space curve in the $(4^{n}-1)$-D space spanned by all Pauli operators $\sigma_{j}$, except for $I$. The diagrammatic approach to studying error dynamics and robust control, known as the Quantum Erroneous Evolution Diagram (QEED)\cite{hai2022universal}, effectively captures all these geometric properties of the noise dynamics. Note that the error operator $\Phi (t)$ is generated by noise operator $V_{i}$ and the interaction picture associated with $U_{0}(t)$, which infers that the same noise might result in different errors depending on $U_{0}(t)$. This intuition reveals the key point of this work: To use geometric formalism to design gate sequences to implement dynamically correcting quantum circuits.

For continuous qubit dynamics, the time-dependent Hamiltonian can be reconstructed from the general curvature of the curve, using the generalized Frenet-Serret formulas in higher-dimensional Euclidean space~\cite{buterakosGeometricalFormalismDynamically2021,hai2022universal}. The simplest example is the resonantly-driven single qubit under the quasi-static dephasing noise, where the driving field amplitude corresponds to the curvature of a curve lying in a two-dimensional plane\cite{zeng2018general}. Instead of analyzing continuous quantum dynamics, this paper generalizes the QEED approach to the dynamics of quantum circuits, where overall quantum evolution is decomposed into discrete layers of gates, so as $\sum_{i}R_{j}^{(i)}(t)\rightarrow \Gamma _{j}^{[l]}$. Here, the symbol $\mathbf{\Gamma }^{[l]}$ represents the discrete version of the erroneous evolution curve generated by $U_{\epsilon}^{[l]}=e^{-i \Phi^{[l]}}$ at the circuit layer $l$. Notice that we have omitted the noise source index $i$ by absorbing the summation over $i$ into the definition of $\Gamma $. 

For a more formal approach, consider a noisy circuit layer indexed $l$ as an ideal gate operator preceded by a local error evolution operator, $U_0^{[l]} e^{-i \mathbf{\Gamma}_{\text{lo}}^{[l]}\cdot \mathbf{\sigma} }$, similar to Eq.~\ref{eq:decom}. 
We can move evolution operators attached to all noisy layers to the beginning of the circuit to analyze error accumulation throughout the whole circuit. Two properties are useful in this context: a chunk of quantum circuits represented by unitary operator $\mathcal{C}$ and the error evolution $e^{-i\Phi}$ follows the commutation rule $e^{-i\Phi}\mathcal{C}\approx \mathcal{C}e^{-i\mathcal{C}^{\dagger}\Phi\mathcal{C}}$, and for weak noise, we can treat errors at different circuit layers as commuting with each other, $e^{-i\Phi_i}e^{-i\Phi_j} \approx e^{-i\Phi_j}e^{-i\Phi_i}$. Thus, the error step vector after this move $\mathbf{\Gamma}^{[l]}$ is connected with local error steps $\mathbf{\Gamma}_{\text{lo}}^{[l]}$ by $\Gamma _{j}^{[l]}=\frac{1}{D}\Tr(\mathcal{C}^{[l]\dagger }\left( \mathbf{\Gamma }_{\text{lo}}^{[l]}\cdot \mathbf{\sigma }\right) \mathcal{C}^{[l]}\sigma_{j})$, and the total error operator: 
\begin{equation}
\Phi_{\text{total}}=\sum_{l=1}^{N_{D}}\sum_{j}\Gamma _{j}^{[l]}\sigma_{j}  
\label{eq:Phitotal}
\end{equation}
Here $N_{D}$ is the circuit depth, $j$ covers all indices of the generalized Pauli matrices, and $\mathcal{C}^{[l]}$ represents the circuit from the 1st to $l$'th layer. Since $\mathcal{C}^{[l]}$ is unitary and thus norm-preserving, error vectors $\mathbf{\Gamma }^{[l]}$ and $\mathbf{\Gamma }_{\text{lo}}^{[l]}$ are connected through a rotation transformation. Eq.~\ref{eq:Phitotal} can be seen as a path formed by $N_{D}$ steps in the QEED, with each step vector being $\mathbf{\Gamma} ^{\lbrack l]}$.

\begin{figure}[tbp]
\includegraphics[width=\columnwidth]{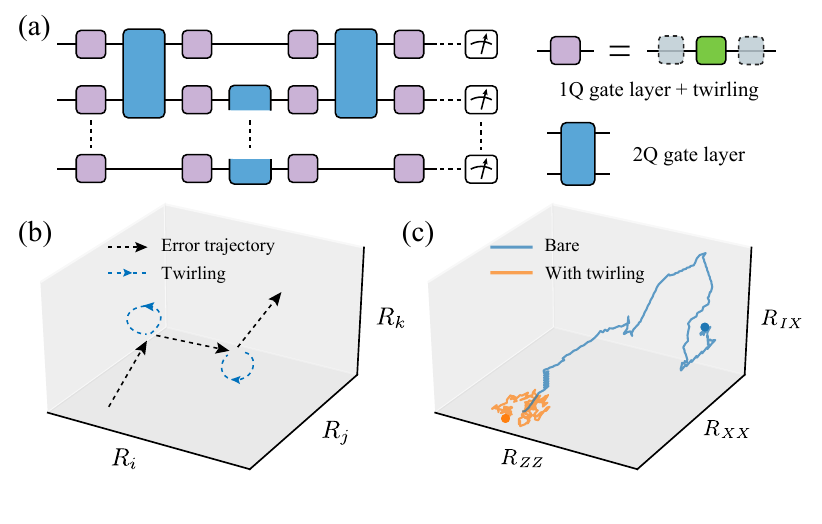} 
\caption{A visualization of a quantum circuit with twirling gates (a) and its corresponding error trajectory (b). Twirling gates are assumed to be error-free and, therefore, correspond to closed loops in the quantum erroneous evolution diagram. (c) shows quantum erroneous diagrams for the circuit errors with and without Pauli twirling gates. Although all Pauli components contain errors, only three are displayed for clarity.}
\label{fig:RC_picture}
\end{figure}

% \subsection{Noise Tailoring With Twirling Gates}
\textit{Enhancing robustness via twirling sequence}.- Previously, we have illustrated the geometric approach of error dynamics at the circuit level. In the following, we will show how to suppress the errors by optimizing circuit robustness with twirling sequences and further improve noise resilience by integrating robust pulse engineering. We introduce the concept of \textquotedblleft hard'' and \textquotedblleft easy'' layers within quantum circuits. Hard layers, typically involving multi-qubit gates, are more noise-prone, whereas easy layers, usually consisting of single-qubit gates, are relatively error-free. A quantum circuit can be seen as a series of alternating easy and hard layers (consecutive easy layers can be combined, and ideal identity layers can be inserted between consecutive hard layers), as shown in Fig.~\ref{fig:RC_picture}(a) with purple and blue boxes. To generate an equivalent circuit, we insert a twirling operator $T$ before a hard layer and a correction operator $T^c$ afterward, ensuring the modified layer performs the same operation, satisfying $T^{c}U_{0}T=U_{0}$. These twirling and correction operators can be combined with easy layers to ensure that they don't increase the overhead of the quantum circuit in terms of both circuit depth and noise.
The twirled noisy hard layer is $T^{c}U_{0}e^{-i\left( \mathbf{\Gamma }_{\text{lo}}\cdot \mathbf{\sigma }\right) }T$. Commuting the error operator and twirling gate, we obtain $%
U_{0}e^{-iT^{\dagger }\left( \mathbf{\Gamma }_{\text{lo}}\cdot \mathbf{\sigma }\right) T} = U_{0}e^{-i \tilde{\mathbf{\Gamma }}_{\text{lo}}\cdot \mathbf{\sigma }}$. Geometrically, this can be seen as a twirling operation $T$ rotating
step error vector $\mathbf{\Gamma }_{\text{lo}}$ into $\tilde{\mathbf{\Gamma }}_{\text{lo}}$, with the process illustrated as the dashed
blue curves in Fig.\ref{fig:RC_picture}(b), modifying the error curves in
the QEED associated with the circuit. Specifically, for Pauli twirling $T=\sigma_{i}$, $\tilde{\Gamma}_{\text{lo}}$ with each of its component $\tilde{\Gamma}_{\text{lo}, i}=\gamma_i\Gamma_{\text{lo}, i}$ where $\gamma_i=1$ if $T$ at this layer commutes with $\sigma_i$ and $-1$ if they anti-commutes. As the layers increase to $N_{D}$, the total error
distance of the twirled circuit becomes $|\tilde{\Phi} _{\text{total}}|=\left|\sum_{i}^{N_{D}}\mathbf{\tilde{\Gamma}} ^{[i]}\right|$ as opposed to the original circuit $|\Phi _{\text{total}}|=\left|\sum_{i}^{N_{D}}\mathbf{\Gamma} ^{[i]}\right|$. Here $\mathbf{\tilde{\Gamma}} ^{[i]}$ is obtained from $\mathbf{\tilde{\Gamma}}_{\text{lo}} ^{[i]}$ by the same mapping as the original circuit, $\tilde{\Gamma} _{j}^{[i]}=\frac{1}{D}\Tr(\mathcal{C}^{[i]\dagger }\left( \tilde{\mathbf{\Gamma }}_{\text{lo}}^{[i]}\cdot \mathbf{\sigma }\right) \mathcal{C}^{[i]}\sigma_{j})$.  %and the one with twirling can be expressed as 
%\begin{equation}
%\begin{array}{c}
%|\Phi _{\text{total}}|^{2}=\sum_{i}^{N_{D}}\left|\mathbf{\Gamma} ^{[i]}\right|^{2} \\ 
%|\tilde{\Phi}_{\text{total}}|^{2}=|\Phi _{\text{total}}|^{2}+2%
%\sum_{i<k}^{N_{D}}\sum_{j}\left(\gamma_{j}^{[i]}\gamma_{j}^{[k]}-1\right)\Gamma_{j}^{[i]}\Gamma_{j}^{[k]}%
%\end{array}%
%\label{Eq_TwirledError}
%\end{equation}%
%, where $\gamma_j^{[i]}=1$ if $T$ at layer $[i]$ commutes with $\sigma_j$ and $-1$ if anti-commutes. 
In principle, we can design a sequence of twirling gates to make the twirled error operator $|\tilde{\Phi}_{\text{total}}|<| \Phi _{\text{total}}|$, so that the errors are suppressed consequently. Fig.~\ref{fig:RC_picture}(c) visually illustrates how a sequence of Pauli twirling gates morphs the error trajectory and makes it closer to the origin. The original two-qubit circuit has 200 layers of noisy CNOT gates and 74 noise-free Hadamard gates. The figure contrasts the error trajectories of the bare circuit and the modified, more robust twirled circuit, clearly highlighting the error reduction attributable to these twirling gates. It is important to note that randomly chosen twirling sequences might actually increase errors, and searching for error-suppressing twirling sequences is computationally hard in general. Various heuristic optimization methods could be applied to search for optimal twirling sequences. For small systems with shallow circuits, a brute-force search is feasible, and a smaller set of twirling gates can be adapted when the error does not span the entire Pauli space, as discussed in~\cite{cai2019constructing}. 

\begin{figure}
    \includegraphics[width=\columnwidth]{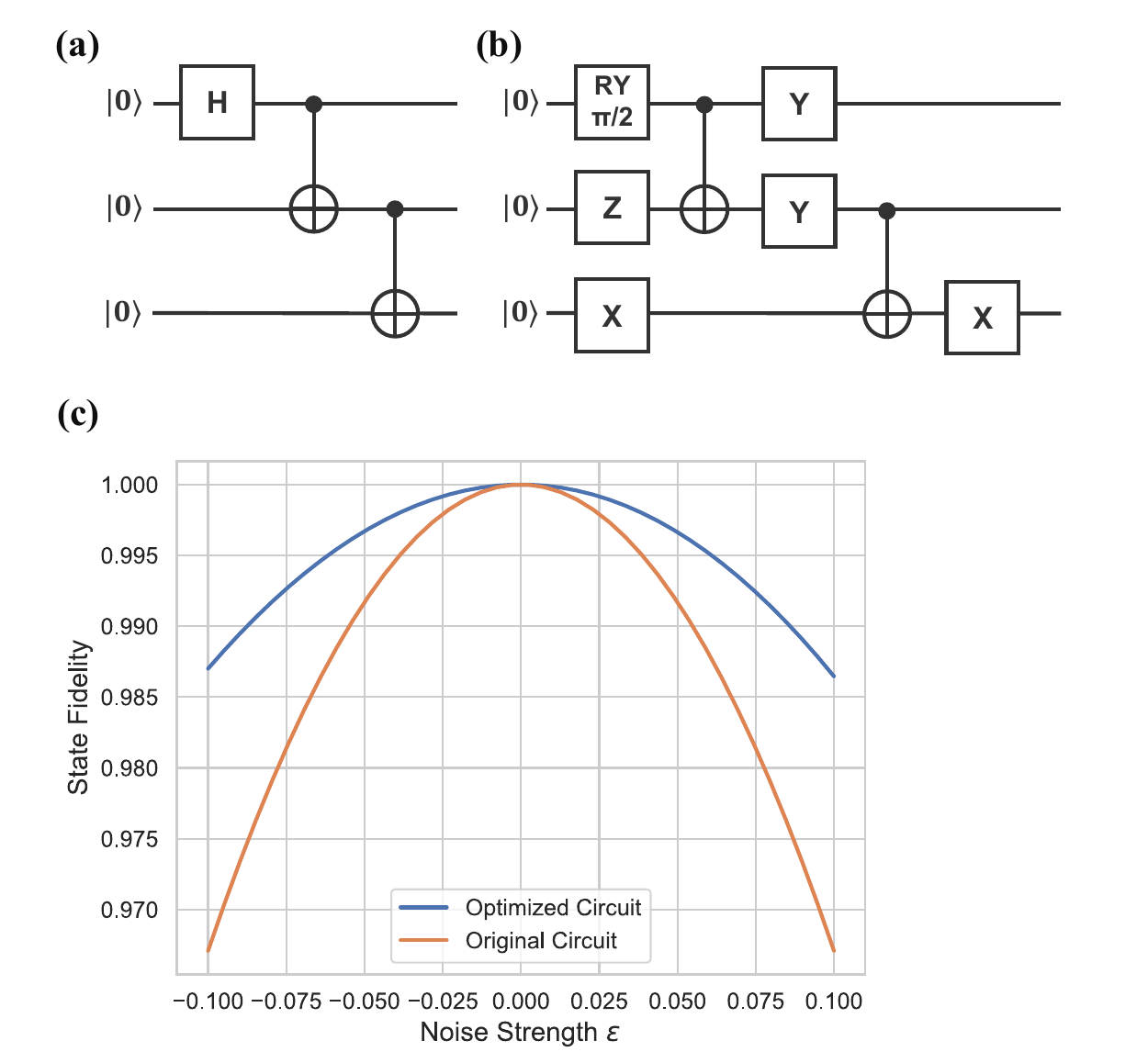}
    \caption{(a) 3-qubit GHZ state preparation circuit. (b) GHZ circuit optimized via twirling gates robust to error type $(Z+Y)I+(Z-I)(Y+Z)$. (c) A comparison of fidelity vs noise strength between two GHZ circuits.}
    \label{fig:GHZ_circuit}
\end{figure}

We use the Greenberger–Horne–Zeilinger(GHZ) state preparation circuit as an example to demonstrate how a standard circuit can be made more noise-resistant by incorporating optimized twirling operations, as depicted in Fig.~\ref{fig:GHZ_circuit}(a) and (b). In the circuit, CNOT gates are attached with an error operator $\Phi =\epsilon \frac{1}{\sqrt{6}}\left(
(Z+Y)I+(Z-I)(Y+Z)\right) $ so that both single- and two-qubit noises are mimicked. Our numerical analysis reveals that the twirled circuit shows remarkably higher noise resistance compared to the original circuit, as shown in
Fig.~\ref{fig:GHZ_circuit}(c). 
%It is worth noting that the twirling gate optimization process is discrete and typically presents significant complexity. In the context of three-qubit GHZ state preparation, where the circuit size remains relatively small, it is feasible to employ a brute-force search approach. However, for larger circuits with higher complexity, more heuristic optimization methods will be needed, and we leave the exploration of circuit optimization for future studies.

\textit{Randomized compiling, random walk, and robust control}.- While the search for twirling sequences that optimally suppress the errors is hard due to noise characterization and the nature of discrete optimization problems, we will revisit randomized compiling (RC) with the QEED analysis and show that averaging random twirling sequences could mitigate errors and, integrated with robust pulses, it can further improve circuit fidelity. If all Pauli matrices, including the identity, have an equal chance of being chosen for $T$, a plus or minus sign may appear at all components of each error step with a 50/50 chance. The error path is turned into a variant of random walk process in the space of generalized Pauli matrices, causing the components of the error operator $\tilde{\Phi}_{\text{total}}$ to follow a Gaussian distribution and the total error distance $|\tilde{\Phi}_{\text{total}}|$ square-root scales with the circuit depth $N_{D}$ as $\langle|\Phi_{\text{total}}|\rangle \sim \sqrt{N_{D}}\Gamma$, which is precisely what an N-dimensional random walk process would predict.  Here $\Gamma$ is the root mean square of the error step lengths, $\Gamma=\sqrt{\frac{1}{N_D}\sum_{i}\left| \mathbf{\Gamma}^{[i]}\right| ^{2}}$. We discuss more on this property in the supplementary.

We now analyze the error after averaging multiple different instances of twirled circuits, focusing on the behavior of one single circuit layer. The noise channel on operator $P$ for one circuit layer is $\mathcal{E}(P) = \mathbb{E}\left(\tilde{U}_\epsilon P \tilde{U}_\epsilon^\dagger\right)$. The expectation operator $\mathbb{E}$ is defined by averaging over all possible twirling gates. Here, we use a super operator instead of a unitary operator, as classical averaging results in a mixed state. The expanded error evolution, $\tilde{U}_\epsilon = \exp\left(-i\sum_{n=1}^{\infty} \frac{1}{n!}\sum_i \epsilon^n \Gamma_i^{n}T\sigma_i T\right)$, includes n-th order errors measured by $\Gamma_i^n$ and twirling operator $T$. Note that here we use the no-bracket upper index of $\Gamma_i^n$ for the error's order in the Magnus expansion rather than for the circuit layer as in previous sections. We perform a Taylor expansion on the noise channel around $\epsilon=0$ to obtain its explicit form approximately. The first orders in the expansion, $\mathcal{E}(P)=\mathcal{E}_\epsilon^{0-3}(P)+\mathcal{E}_\epsilon^{4}(P)$, are calculated as
\begin{equation}
\label{eq:noisechannelexp}
    \begin{split}
    &\mathcal{E}_\epsilon^{0-3}(P) = P - 2\sum_{\{\sigma_i,P\}=0} \Gamma_{i}^{1}\left( \Gamma^{1}_i +\Gamma^{2}_i\right) P \\
        &\mathcal{E}_\epsilon^{4}(P)=\sum_{\{\sigma{i},P\}=0} \left( \frac{2}{3}\Gamma^{1}_i \left(\left(\Gamma^{1}_i\right)^3-\Gamma^{3}_i\right)-\frac{1}{2}\left(\Gamma^{2}_i\right)^2 \right)P\\ &+\sum_{\substack{\{\sigma_{\{i,j\}},P\}=0\\
                                                [\sigma_i, \sigma_j]=0\\
                                                i\neq j
                                                }
                                    }2\left(\Gamma^{1}_i \Gamma^{1}_j\right)^2 P%\\
    \end{split}
\end{equation}
We will discuss in more detail how to derive the formula in the supplementary. %\ref{app:noisech}. 

Deriving the analytical form of the resulting Pauli noise channel from RC as in \ref{eq:noisechannelexp} is crucial not only for assessing errors but also for estimating their inverse operations. This enables the implementation of quantum error mitigation techniques to restore the ideal state through data post-processing. Importantly, this process necessitates noise characterization to gather information on $\Gamma_i^{n}$.

\begin{figure}[tbp]
    \includegraphics[width=\columnwidth]{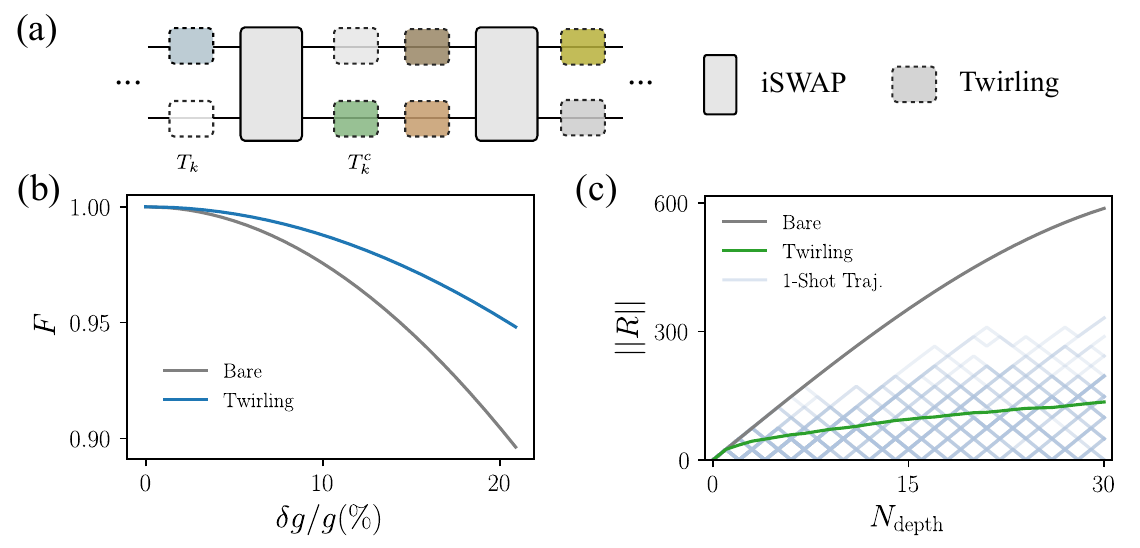}
    \caption{(a) A visualization of an iSWAP circuit with twirling gates added. (b) State transfer fidelity of one single layer in the iSWAP circuit. Bare circuit and randomized compiled circuit are compared. (c) Error distances of the iSWAP circuit as the circuit depth grows for the cases of the bare circuit, the circuit with randomly applied twirling gates, and the averaged randomized compiled circuits.}
    \label{fig:iSWAP}
\end{figure}

\begin{figure*}[tbp]
    \includegraphics[width=2\columnwidth]{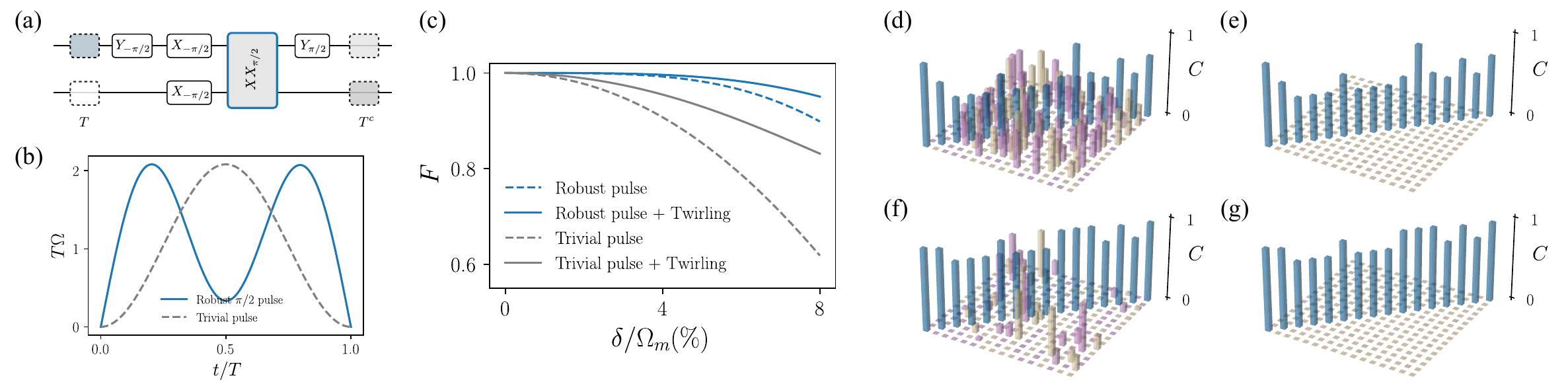}
    \caption{(a)A visualization of the CNOT circuit. The circuit is composed of an $XX(\pi/2)$ gate and several single-qubit Clifford gates. Random twirling gates $T_k$ and the corresponding compensate gates $T_k^c$ are represented with dotted line boxes. (b) The dephasing-robust optimized control pulse (blue) and the trivial cosine pulse (grey) for a $XX(\frac{\pi}{2})$ gate. (c) State transfer fidelity of one single layer in the circuit of CNOT that transfers $|01\rangle$ to $|10\rangle$ and $1/\sqrt{2}(|0\rangle+|1\rangle)|0\rangle$ to $1/\sqrt{2}(|00\rangle+|11\rangle)$, respectively. The x-axis is the noise strength. The bare circuit and randomized compiled circuit are compared, and the circuit with gates implemented with robust control pulse waveform in (b) is compared. (d)-(g) PTM of the noise channel in the CNOT circuit under different conditions, where the diagonal elements are marked in blue; negative and non-negative off-diagonal elements are marked in plum and wheat, respectively. (d) Bare circuit with the trivial pulse, (e) Circuit with randomized compiling and trivial pulse, (f) Bare circuit with a robust pulse, and (g) Circuit with both randomized compiling and a robust pulse. The figures highlight how randomized compiling eliminates off-diagonal PTM terms, and robust pulses enhance the values of diagonal elements towards unity. }
    \label{fig:CNOT}
\end{figure*}

% \section{Numerical Results}
While RC already enhances circuit fidelity by averaging out coherent part of the error, optimizing gate pulse implementation to make it first-order error-robust as $\Gamma_i^{1}=0$—where the gate-associated QEED forms a closed curve—can substantially reduce many terms in the noise channel.
We now numerically show how RC improves circuit fidelity, and incorporating robust control techniques takes the fidelity improvement even further. Consider two specific and simple circuits for demonstration purposes: a cycling iSWAP circuit and a composed CNOT circuit. The iSWAP gate can be generated by a Hamiltonian $g(XX + YY)$ and exposed to a Heisenburg-style $\delta g(XX+YY+ZZ)$ noise term, as shown in Fig.\ref{fig:iSWAP}. The CNOT circuit we use consists of a two-qubit gate $XX(\pi/2)$ and several single-qubit gates, as shown in Fig.~\ref{fig:CNOT}. We assume these single-qubit gates to be noiseless, with the Hamiltonian generating the $XX$ gate given by $\frac{1}{2} \Omega(t) XX$ and noise term $\delta(IZ-ZI+\frac{1}{2}ZZ)$. $\Omega(t)$ is the control pulse waveform. Note that since $XX$ does not commute with either $ZI$ or $IZ$, the Hamiltonian for the $XX$ gate can potentially dynamically correct errors caused by these noise sources while leaving $ZZ$ noise term unaltered. On the other hand, for the iSWAP case, the noise is completely uncorrectable since the noise term commutes with all terms in the Hamiltonian. We quantify the error accumulated in the circuit through the error distance $||\mathbf{R}||$. For the iSWAP case, the error distance forms a straight line as the circuit goes deeper. Such a situation corresponds to a worst-case scenario for which the randomized compiling technique can significantly improve fidelity.

In Fig.~\ref{fig:iSWAP}(c), we show that with the application of twirling gates, the error distance exhibits a one-dimensional random walk pattern, leading to significantly slower error accumulation compared to the raw iSWAP circuit. When multiple instances with randomly sampled twirling sequences are averaged, as in the case of RC, the error distance increases proportionally to the square root of the circuit depth, consistent with the characteristics of a random walk process.
We also demonstrate fidelity improvements in randomized compiled circuits at a single circuit layer. 
Fig.~\ref{fig:iSWAP}(b) and Fig.~\ref{fig:CNOT}(c) compare state transfer fidelity under various noise strengths, both with and without twirling gates.
The figures clearly show that incorporating RC significantly boosts fidelity in both CNOT and iSWAP circuits.

Because part of the noise source in the CNOT circuit is potentially correctable, we also incorporated pulse-level optimal quantum control so that several components in $\mathbf{\Gamma}^{1}$ vanish. 
The corresponding $\pi/2$ pulses with an optimized shape obtained in~\cite{hai2022universal} and a trivial cosine shape are shown in Fig.~\ref{fig:CNOT}(b). 
The optimal control pulse exhibits robustness against IZ and ZI noises. 
It shows clearly in the figure that the fidelity is further improved on top of the already-existing improvement from random twirling gates, which demonstrates that gate-level robustness from quantum control techniques can complement circuit compiling level robustness, and the advantage gained from robust control is separated with RC.
%\begin{figure}
    %\includegraphics[width=\columnwidth]{image}
%    \includegraphics[width=0.6\columnwidth]{pulse.pdf}
%    \caption{The dephasing-robust optimized control pulse (blue) and the trivial cosine pulse (grey) for a $XX(\frac{\pi}{2})$ gate.}
    %\includegraphics[width=3cm]{example-grid-100x100pt}
%    \label{fig:pulse}
%\end{figure}

We further analyzed the effects of RC and robust control by visualizing the noise channel as a Pauli transfer matrix (PTM). 
Figures (d) and (f) in Fig.\ref{fig:CNOT} represent the circuit without twirling gates, while figures (e) and (g) illustrate the results with RC applied. 
RC eliminates all cross terms in the noise channel, converting it into a Pauli channel. 
This is reflected in the PTM as a diagonal matrix, with diagonal elements defined by Eq.\ref{eq:noisechannelexp}. 
We also incorporated robust quantum control techniques in the two bottom figures. 
These show that circuits implemented with robust control pulses have PTM diagonal elements closer to 1, indicating enhanced fidelity due to the robust control pulses.
%\begin{figure}
%\includegraphics[width=\columnwidth]{PTM_new.pdf}
  % \includegraphics[width=\columnwidth]{PTM.pdf}
  %\includesvg[width=\columnwidth]{CNOT_robust_2d.svg}
 %   \caption{PTM of the noise channel in the CNOT circuit under different conditions, where the diagonal elements are marked in blue; negative and non-negative off-diagonal elements are marked in plum and wheat, respectively. (a) Bare circuit with the trivial pulse, (b) Circuit with randomized compiling and trivial pulse, (c) Bare circuit with a robust pulse, and (d) Circuit with both randomized compiling and a robust pulse. The figures highlight how randomized compiling eliminates off-diagonal PTM terms, and robust pulses enhance the values of diagonal elements towards unity.}
  %  \label{fig:ptm}
%\end{figure}

In conclusion, we have expanded the geometric framework of noisy quantum dynamics to the circuit level, mapping noise-induced error accumulation to a trajectory in QEED. This framework enables the optimization of quantum circuits for enhanced robustness. We demonstrated that integrating twirling gates transforms circuits into equivalents with different error susceptibilities, offering the potential for robustness optimization. Random twirling gates reshape the error trajectory into a random walk process, significantly reducing worst-case error. Averaging multiple equivalent circuits eliminates noise coherence in the channel, converting it into a Pauli noise channel. Parameters of this noise channel are determined from the error distance of individual quantum gates, also depicted using geometric formalism. Lastly, we showed that incorporating robust quantum control techniques further improves circuit fidelity, as illustrated through various circuit examples.

\begin{acknowledgments}
This work was supported by the Key-Area Research and Development Program of Guang-Dong Province (Grant No. 2018B030326001), the National Natural Science Foundation of China (U1801661), the Guangdong Innovative and Entrepreneurial Research Team Program (2016ZT06D348), the Guangdong Provincial Key Laboratory (Grant No.2019B121203002), the Natural Science Foundation of Guangdong Province (2017B030308003), and the Science, Technology and Innovation Commission of Shenzhen Municipality (JCYJ20170412152620376, KYTDPT20181011104202253), and the NSF of Beijing (Grants No. Z190012), Shenzhen Science and Technology Program (KQTD20200820113010023).
\end{acknowledgments}

%\appendix

%\section{Appendixes}

% The \nocite command causes all entries in a bibliography to be printed out
% whether or not they are actually referenced in the text. This is an appropriate
% for the sample file to show the different styles of references, but authors
% most likely will not want to use it.
%\nocite{*}

\bibliography{apssamp}% Produces the bibliography via BibTeX.

%apsrev4-2.bst 2019-01-14 (MD) hand-edited version of apsrev4-1.bst
%Control: key (0)
%Control: author (72) initials jnrlst
%Control: editor formatted (1) identically to author
%Control: production of article title (-1) disabled
%Control: page (0) single
%Control: year (1) truncated
%Control: production of eprint (0) enabled
\begin{thebibliography}{35}%
\makeatletter
\providecommand \@ifxundefined [1]{%
 \@ifx{#1\undefined}
}%
\providecommand \@ifnum [1]{%
 \ifnum #1\expandafter \@firstoftwo
 \else \expandafter \@secondoftwo
 \fi
}%
\providecommand \@ifx [1]{%
 \ifx #1\expandafter \@firstoftwo
 \else \expandafter \@secondoftwo
 \fi
}%
\providecommand \natexlab [1]{#1}%
\providecommand \enquote  [1]{``#1''}%
\providecommand \bibnamefont  [1]{#1}%
\providecommand \bibfnamefont [1]{#1}%
\providecommand \citenamefont [1]{#1}%
\providecommand \href@noop [0]{\@secondoftwo}%
\providecommand \href [0]{\begingroup \@sanitize@url \@href}%
\providecommand \@href[1]{\@@startlink{#1}\@@href}%
\providecommand \@@href[1]{\endgroup#1\@@endlink}%
\providecommand \@sanitize@url [0]{\catcode `\\12\catcode `\$12\catcode `\&12\catcode `\#12\catcode `\^12\catcode `\_12\catcode `\%12\relax}%
\providecommand \@@startlink[1]{}%
\providecommand \@@endlink[0]{}%
\providecommand \url  [0]{\begingroup\@sanitize@url \@url }%
\providecommand \@url [1]{\endgroup\@href {#1}{\urlprefix }}%
\providecommand \urlprefix  [0]{URL }%
\providecommand \Eprint [0]{\href }%
\providecommand \doibase [0]{https://doi.org/}%
\providecommand \selectlanguage [0]{\@gobble}%
\providecommand \bibinfo  [0]{\@secondoftwo}%
\providecommand \bibfield  [0]{\@secondoftwo}%
\providecommand \translation [1]{[#1]}%
\providecommand \BibitemOpen [0]{}%
\providecommand \bibitemStop [0]{}%
\providecommand \bibitemNoStop [0]{.\EOS\space}%
\providecommand \EOS [0]{\spacefactor3000\relax}%
\providecommand \BibitemShut  [1]{\csname bibitem#1\endcsname}%
\let\auto@bib@innerbib\@empty
%</preamble>
\bibitem [{\citenamefont {Arute}\ \emph {et~al.}(2019)\citenamefont {Arute}, \citenamefont {Arya}, \citenamefont {Babbush}, \citenamefont {Bacon}, \citenamefont {Bardin}, \citenamefont {Barends}, \citenamefont {Biswas}, \citenamefont {Boixo}, \citenamefont {Brandao}, \citenamefont {Buell} \emph {et~al.}}]{arute2019quantum}%
  \BibitemOpen
  \bibfield  {author} {\bibinfo {author} {\bibfnamefont {F.}~\bibnamefont {Arute}}, \bibinfo {author} {\bibfnamefont {K.}~\bibnamefont {Arya}}, \bibinfo {author} {\bibfnamefont {R.}~\bibnamefont {Babbush}}, \bibinfo {author} {\bibfnamefont {D.}~\bibnamefont {Bacon}}, \bibinfo {author} {\bibfnamefont {J.~C.}\ \bibnamefont {Bardin}}, \bibinfo {author} {\bibfnamefont {R.}~\bibnamefont {Barends}}, \bibinfo {author} {\bibfnamefont {R.}~\bibnamefont {Biswas}}, \bibinfo {author} {\bibfnamefont {S.}~\bibnamefont {Boixo}}, \bibinfo {author} {\bibfnamefont {F.~G.}\ \bibnamefont {Brandao}}, \bibinfo {author} {\bibfnamefont {D.~A.}\ \bibnamefont {Buell}}, \emph {et~al.},\ }\href@noop {} {\bibfield  {journal} {\bibinfo  {journal} {Nature}\ }\textbf {\bibinfo {volume} {574}},\ \bibinfo {pages} {505} (\bibinfo {year} {2019})}\BibitemShut {NoStop}%
\bibitem [{\citenamefont {Wu}\ \emph {et~al.}(2021)\citenamefont {Wu}, \citenamefont {Bao}, \citenamefont {Cao}, \citenamefont {Chen}, \citenamefont {Chen}, \citenamefont {Chen}, \citenamefont {Chung}, \citenamefont {Deng}, \citenamefont {Du}, \citenamefont {Fan} \emph {et~al.}}]{wu2021strong}%
  \BibitemOpen
  \bibfield  {author} {\bibinfo {author} {\bibfnamefont {Y.}~\bibnamefont {Wu}}, \bibinfo {author} {\bibfnamefont {W.-S.}\ \bibnamefont {Bao}}, \bibinfo {author} {\bibfnamefont {S.}~\bibnamefont {Cao}}, \bibinfo {author} {\bibfnamefont {F.}~\bibnamefont {Chen}}, \bibinfo {author} {\bibfnamefont {M.-C.}\ \bibnamefont {Chen}}, \bibinfo {author} {\bibfnamefont {X.}~\bibnamefont {Chen}}, \bibinfo {author} {\bibfnamefont {T.-H.}\ \bibnamefont {Chung}}, \bibinfo {author} {\bibfnamefont {H.}~\bibnamefont {Deng}}, \bibinfo {author} {\bibfnamefont {Y.}~\bibnamefont {Du}}, \bibinfo {author} {\bibfnamefont {D.}~\bibnamefont {Fan}}, \emph {et~al.},\ }\href@noop {} {\bibfield  {journal} {\bibinfo  {journal} {Phys. Rev. Lett.}\ }\textbf {\bibinfo {volume} {127}},\ \bibinfo {pages} {180501} (\bibinfo {year} {2021})}\BibitemShut {NoStop}%
\bibitem [{\citenamefont {Daley}\ \emph {et~al.}(2022)\citenamefont {Daley}, \citenamefont {Bloch}, \citenamefont {Kokail}, \citenamefont {Flannigan}, \citenamefont {Pearson}, \citenamefont {Troyer},\ and\ \citenamefont {Zoller}}]{daley2022practical}%
  \BibitemOpen
  \bibfield  {author} {\bibinfo {author} {\bibfnamefont {A.~J.}\ \bibnamefont {Daley}}, \bibinfo {author} {\bibfnamefont {I.}~\bibnamefont {Bloch}}, \bibinfo {author} {\bibfnamefont {C.}~\bibnamefont {Kokail}}, \bibinfo {author} {\bibfnamefont {S.}~\bibnamefont {Flannigan}}, \bibinfo {author} {\bibfnamefont {N.}~\bibnamefont {Pearson}}, \bibinfo {author} {\bibfnamefont {M.}~\bibnamefont {Troyer}},\ and\ \bibinfo {author} {\bibfnamefont {P.}~\bibnamefont {Zoller}},\ }\href@noop {} {\bibfield  {journal} {\bibinfo  {journal} {Nature}\ }\textbf {\bibinfo {volume} {607}},\ \bibinfo {pages} {667} (\bibinfo {year} {2022})}\BibitemShut {NoStop}%
\bibitem [{\citenamefont {Bentley}\ \emph {et~al.}(2020)\citenamefont {Bentley}, \citenamefont {Ball}, \citenamefont {Biercuk}, \citenamefont {Carvalho}, \citenamefont {Hush},\ and\ \citenamefont {Slatyer}}]{bentleyNumericOptimizationConfigurable2020}%
  \BibitemOpen
  \bibfield  {author} {\bibinfo {author} {\bibfnamefont {C.~D.~B.}\ \bibnamefont {Bentley}}, \bibinfo {author} {\bibfnamefont {H.}~\bibnamefont {Ball}}, \bibinfo {author} {\bibfnamefont {M.~J.}\ \bibnamefont {Biercuk}}, \bibinfo {author} {\bibfnamefont {A.~R.~R.}\ \bibnamefont {Carvalho}}, \bibinfo {author} {\bibfnamefont {M.~R.}\ \bibnamefont {Hush}},\ and\ \bibinfo {author} {\bibfnamefont {H.~J.}\ \bibnamefont {Slatyer}},\ }\href@noop {} {\bibfield  {journal} {\bibinfo  {journal} {arXiv:2005.00366}\ } (\bibinfo {year} {2020})},\ \Eprint {https://arxiv.org/abs/2005.00366} {2005.00366} \BibitemShut {NoStop}%
\bibitem [{\citenamefont {Cody~Jones}\ \emph {et~al.}(2012)\citenamefont {Cody~Jones}, \citenamefont {Ladd},\ and\ \citenamefont {Fong}}]{codyjonesDynamicalDecouplingQubit2012}%
  \BibitemOpen
  \bibfield  {author} {\bibinfo {author} {\bibfnamefont {N.}~\bibnamefont {Cody~Jones}}, \bibinfo {author} {\bibfnamefont {T.~D.}\ \bibnamefont {Ladd}},\ and\ \bibinfo {author} {\bibfnamefont {B.~H.}\ \bibnamefont {Fong}},\ }\href {https://doi.org/10.1088/1367-2630/14/9/093045} {\bibfield  {journal} {\bibinfo  {journal} {New J. Phys.}\ }\textbf {\bibinfo {volume} {14}},\ \bibinfo {pages} {093045} (\bibinfo {year} {2012})}\BibitemShut {NoStop}%
\bibitem [{\citenamefont {Edmunds}\ \emph {et~al.}(2020)\citenamefont {Edmunds}, \citenamefont {Hempel}, \citenamefont {Harris}, \citenamefont {Frey}, \citenamefont {Stace},\ and\ \citenamefont {Biercuk}}]{edmundsDynamicallyCorrectedGates2020}%
  \BibitemOpen
  \bibfield  {author} {\bibinfo {author} {\bibfnamefont {C.~L.}\ \bibnamefont {Edmunds}}, \bibinfo {author} {\bibfnamefont {C.}~\bibnamefont {Hempel}}, \bibinfo {author} {\bibfnamefont {R.~J.}\ \bibnamefont {Harris}}, \bibinfo {author} {\bibfnamefont {V.}~\bibnamefont {Frey}}, \bibinfo {author} {\bibfnamefont {T.~M.}\ \bibnamefont {Stace}},\ and\ \bibinfo {author} {\bibfnamefont {M.~J.}\ \bibnamefont {Biercuk}},\ }\href {https://doi.org/10.1103/PhysRevResearch.2.013156} {\bibfield  {journal} {\bibinfo  {journal} {Phys. Rev. Research}\ }\textbf {\bibinfo {volume} {2}},\ \bibinfo {pages} {013156} (\bibinfo {year} {2020})}\BibitemShut {NoStop}%
\bibitem [{\citenamefont {Green}\ \emph {et~al.}(2013)\citenamefont {Green}, \citenamefont {Sastrawan}, \citenamefont {Uys},\ and\ \citenamefont {Biercuk}}]{greenArbitraryQuantumControl2013}%
  \BibitemOpen
  \bibfield  {author} {\bibinfo {author} {\bibfnamefont {T.~J.}\ \bibnamefont {Green}}, \bibinfo {author} {\bibfnamefont {J.}~\bibnamefont {Sastrawan}}, \bibinfo {author} {\bibfnamefont {H.}~\bibnamefont {Uys}},\ and\ \bibinfo {author} {\bibfnamefont {M.~J.}\ \bibnamefont {Biercuk}},\ }\href {https://doi.org/10.1088/1367-2630/15/9/095004} {\bibfield  {journal} {\bibinfo  {journal} {New J. Phys.}\ }\textbf {\bibinfo {volume} {15}},\ \bibinfo {pages} {095004} (\bibinfo {year} {2013})}\BibitemShut {NoStop}%
\bibitem [{\citenamefont {Hansen}\ \emph {et~al.}(2021)\citenamefont {Hansen}, \citenamefont {Seedhouse}, \citenamefont {Saraiva}, \citenamefont {Laucht}, \citenamefont {Dzurak},\ and\ \citenamefont {Yang}}]{hansenPulseEngineeringGlobal2021}%
  \BibitemOpen
  \bibfield  {author} {\bibinfo {author} {\bibfnamefont {I.}~\bibnamefont {Hansen}}, \bibinfo {author} {\bibfnamefont {A.~E.}\ \bibnamefont {Seedhouse}}, \bibinfo {author} {\bibfnamefont {A.}~\bibnamefont {Saraiva}}, \bibinfo {author} {\bibfnamefont {A.}~\bibnamefont {Laucht}}, \bibinfo {author} {\bibfnamefont {A.~S.}\ \bibnamefont {Dzurak}},\ and\ \bibinfo {author} {\bibfnamefont {C.~H.}\ \bibnamefont {Yang}},\ }\href {https://doi.org/10.1103/PhysRevA.104.062415} {\bibfield  {journal} {\bibinfo  {journal} {Phys. Rev. A}\ }\textbf {\bibinfo {volume} {104}},\ \bibinfo {pages} {062415} (\bibinfo {year} {2021})}\BibitemShut {NoStop}%
\bibitem [{\citenamefont {Kabytayev}\ \emph {et~al.}(2014)\citenamefont {Kabytayev}, \citenamefont {Green}, \citenamefont {Khodjasteh}, \citenamefont {Biercuk}, \citenamefont {Viola},\ and\ \citenamefont {Brown}}]{kabytayevRobustnessCompositePulses2014}%
  \BibitemOpen
  \bibfield  {author} {\bibinfo {author} {\bibfnamefont {C.}~\bibnamefont {Kabytayev}}, \bibinfo {author} {\bibfnamefont {T.~J.}\ \bibnamefont {Green}}, \bibinfo {author} {\bibfnamefont {K.}~\bibnamefont {Khodjasteh}}, \bibinfo {author} {\bibfnamefont {M.~J.}\ \bibnamefont {Biercuk}}, \bibinfo {author} {\bibfnamefont {L.}~\bibnamefont {Viola}},\ and\ \bibinfo {author} {\bibfnamefont {K.~R.}\ \bibnamefont {Brown}},\ }\href {https://doi.org/10.1103/PhysRevA.90.012316} {\bibfield  {journal} {\bibinfo  {journal} {Phys. Rev. A}\ }\textbf {\bibinfo {volume} {90}},\ \bibinfo {pages} {012316} (\bibinfo {year} {2014})}\BibitemShut {NoStop}%
\bibitem [{\citenamefont {Khodjasteh}\ \emph {et~al.}(2013)\citenamefont {Khodjasteh}, \citenamefont {Sastrawan}, \citenamefont {Hayes}, \citenamefont {Green}, \citenamefont {Biercuk},\ and\ \citenamefont {Viola}}]{khodjastehDesigningPracticalHighfidelity2013}%
  \BibitemOpen
  \bibfield  {author} {\bibinfo {author} {\bibfnamefont {K.}~\bibnamefont {Khodjasteh}}, \bibinfo {author} {\bibfnamefont {J.}~\bibnamefont {Sastrawan}}, \bibinfo {author} {\bibfnamefont {D.}~\bibnamefont {Hayes}}, \bibinfo {author} {\bibfnamefont {T.~J.}\ \bibnamefont {Green}}, \bibinfo {author} {\bibfnamefont {M.~J.}\ \bibnamefont {Biercuk}},\ and\ \bibinfo {author} {\bibfnamefont {L.}~\bibnamefont {Viola}},\ }\href {https://doi.org/10.1038/ncomms3045} {\bibfield  {journal} {\bibinfo  {journal} {Nat Commun}\ }\textbf {\bibinfo {volume} {4}},\ \bibinfo {pages} {2045} (\bibinfo {year} {2013})}\BibitemShut {NoStop}%
\bibitem [{\citenamefont {Khodjasteh}\ \emph {et~al.}(2010)\citenamefont {Khodjasteh}, \citenamefont {Lidar},\ and\ \citenamefont {Viola}}]{khodjastehArbitrarilyAccurateDynamical2010}%
  \BibitemOpen
  \bibfield  {author} {\bibinfo {author} {\bibfnamefont {K.}~\bibnamefont {Khodjasteh}}, \bibinfo {author} {\bibfnamefont {D.~A.}\ \bibnamefont {Lidar}},\ and\ \bibinfo {author} {\bibfnamefont {L.}~\bibnamefont {Viola}},\ }\href {https://doi.org/10.1103/PhysRevLett.104.090501} {\bibfield  {journal} {\bibinfo  {journal} {Phys. Rev. Lett.}\ }\textbf {\bibinfo {volume} {104}},\ \bibinfo {pages} {090501} (\bibinfo {year} {2010})}\BibitemShut {NoStop}%
\bibitem [{\citenamefont {Khodjasteh}\ and\ \citenamefont {Viola}(2009)}]{khodjastehDynamicallyErrorCorrectedGates2009}%
  \BibitemOpen
  \bibfield  {author} {\bibinfo {author} {\bibfnamefont {K.}~\bibnamefont {Khodjasteh}}\ and\ \bibinfo {author} {\bibfnamefont {L.}~\bibnamefont {Viola}},\ }\href {https://doi.org/10.1103/PhysRevLett.102.080501} {\bibfield  {journal} {\bibinfo  {journal} {Phys. Rev. Lett.}\ }\textbf {\bibinfo {volume} {102}},\ \bibinfo {pages} {080501} (\bibinfo {year} {2009})}\BibitemShut {NoStop}%
\bibitem [{\citenamefont {Khodjasteh}\ and\ \citenamefont {Lidar}(2005)}]{khodjastehFaultTolerantQuantumDynamical2005}%
  \BibitemOpen
  \bibfield  {author} {\bibinfo {author} {\bibfnamefont {K.}~\bibnamefont {Khodjasteh}}\ and\ \bibinfo {author} {\bibfnamefont {D.~A.}\ \bibnamefont {Lidar}},\ }\href {https://doi.org/10.1103/PhysRevLett.95.180501} {\bibfield  {journal} {\bibinfo  {journal} {Phys. Rev. Lett.}\ }\textbf {\bibinfo {volume} {95}},\ \bibinfo {pages} {180501} (\bibinfo {year} {2005})}\BibitemShut {NoStop}%
\bibitem [{\citenamefont {Sauvage}\ and\ \citenamefont {Mintert}(2022)}]{sauvageOptimalControlFamilies2022}%
  \BibitemOpen
  \bibfield  {author} {\bibinfo {author} {\bibfnamefont {F.}~\bibnamefont {Sauvage}}\ and\ \bibinfo {author} {\bibfnamefont {F.}~\bibnamefont {Mintert}},\ }\href {https://doi.org/10.1103/PhysRevLett.129.050507} {\bibfield  {journal} {\bibinfo  {journal} {Phys. Rev. Lett.}\ }\textbf {\bibinfo {volume} {129}},\ \bibinfo {pages} {050507} (\bibinfo {year} {2022})}\BibitemShut {NoStop}%
\bibitem [{\citenamefont {Veldhorst}\ \emph {et~al.}(2015)\citenamefont {Veldhorst}, \citenamefont {Yang}, \citenamefont {Hwang}, \citenamefont {Huang}, \citenamefont {Dehollain}, \citenamefont {Muhonen}, \citenamefont {Simmons}, \citenamefont {Laucht}, \citenamefont {Hudson}, \citenamefont {Itoh} \emph {et~al.}}]{veldhorst2015two}%
  \BibitemOpen
  \bibfield  {author} {\bibinfo {author} {\bibfnamefont {M.}~\bibnamefont {Veldhorst}}, \bibinfo {author} {\bibfnamefont {C.}~\bibnamefont {Yang}}, \bibinfo {author} {\bibfnamefont {J.}~\bibnamefont {Hwang}}, \bibinfo {author} {\bibfnamefont {W.}~\bibnamefont {Huang}}, \bibinfo {author} {\bibfnamefont {J.}~\bibnamefont {Dehollain}}, \bibinfo {author} {\bibfnamefont {J.}~\bibnamefont {Muhonen}}, \bibinfo {author} {\bibfnamefont {S.}~\bibnamefont {Simmons}}, \bibinfo {author} {\bibfnamefont {A.}~\bibnamefont {Laucht}}, \bibinfo {author} {\bibfnamefont {F.}~\bibnamefont {Hudson}}, \bibinfo {author} {\bibfnamefont {K.~M.}\ \bibnamefont {Itoh}}, \emph {et~al.},\ }\href@noop {} {\bibfield  {journal} {\bibinfo  {journal} {Nature}\ }\textbf {\bibinfo {volume} {526}},\ \bibinfo {pages} {410} (\bibinfo {year} {2015})}\BibitemShut {NoStop}%
\bibitem [{\citenamefont {Huang}\ \emph {et~al.}(2019)\citenamefont {Huang}, \citenamefont {Yang}, \citenamefont {Chan}, \citenamefont {Tanttu}, \citenamefont {Hensen}, \citenamefont {Leon}, \citenamefont {Fogarty}, \citenamefont {Hwang}, \citenamefont {Hudson}, \citenamefont {Itoh} \emph {et~al.}}]{huang2019fidelity}%
  \BibitemOpen
  \bibfield  {author} {\bibinfo {author} {\bibfnamefont {W.}~\bibnamefont {Huang}}, \bibinfo {author} {\bibfnamefont {C.}~\bibnamefont {Yang}}, \bibinfo {author} {\bibfnamefont {K.}~\bibnamefont {Chan}}, \bibinfo {author} {\bibfnamefont {T.}~\bibnamefont {Tanttu}}, \bibinfo {author} {\bibfnamefont {B.}~\bibnamefont {Hensen}}, \bibinfo {author} {\bibfnamefont {R.}~\bibnamefont {Leon}}, \bibinfo {author} {\bibfnamefont {M.}~\bibnamefont {Fogarty}}, \bibinfo {author} {\bibfnamefont {J.}~\bibnamefont {Hwang}}, \bibinfo {author} {\bibfnamefont {F.}~\bibnamefont {Hudson}}, \bibinfo {author} {\bibfnamefont {K.~M.}\ \bibnamefont {Itoh}}, \emph {et~al.},\ }\href@noop {} {\bibfield  {journal} {\bibinfo  {journal} {Nature}\ }\textbf {\bibinfo {volume} {569}},\ \bibinfo {pages} {532} (\bibinfo {year} {2019})}\BibitemShut {NoStop}%
\bibitem [{\citenamefont {Xue}\ \emph {et~al.}(2022)\citenamefont {Xue}, \citenamefont {Russ}, \citenamefont {Samkharadze}, \citenamefont {Undseth}, \citenamefont {Sammak}, \citenamefont {Scappucci},\ and\ \citenamefont {Vandersypen}}]{xue2022quantum}%
  \BibitemOpen
  \bibfield  {author} {\bibinfo {author} {\bibfnamefont {X.}~\bibnamefont {Xue}}, \bibinfo {author} {\bibfnamefont {M.}~\bibnamefont {Russ}}, \bibinfo {author} {\bibfnamefont {N.}~\bibnamefont {Samkharadze}}, \bibinfo {author} {\bibfnamefont {B.}~\bibnamefont {Undseth}}, \bibinfo {author} {\bibfnamefont {A.}~\bibnamefont {Sammak}}, \bibinfo {author} {\bibfnamefont {G.}~\bibnamefont {Scappucci}},\ and\ \bibinfo {author} {\bibfnamefont {L.~M.}\ \bibnamefont {Vandersypen}},\ }\href@noop {} {\bibfield  {journal} {\bibinfo  {journal} {Nature}\ }\textbf {\bibinfo {volume} {601}},\ \bibinfo {pages} {343} (\bibinfo {year} {2022})}\BibitemShut {NoStop}%
\bibitem [{\citenamefont {Mills}\ \emph {et~al.}(2022)\citenamefont {Mills}, \citenamefont {Guinn}, \citenamefont {Gullans}, \citenamefont {Sigillito}, \citenamefont {Feldman}, \citenamefont {Nielsen},\ and\ \citenamefont {Petta}}]{mills2022two}%
  \BibitemOpen
  \bibfield  {author} {\bibinfo {author} {\bibfnamefont {A.~R.}\ \bibnamefont {Mills}}, \bibinfo {author} {\bibfnamefont {C.~R.}\ \bibnamefont {Guinn}}, \bibinfo {author} {\bibfnamefont {M.~J.}\ \bibnamefont {Gullans}}, \bibinfo {author} {\bibfnamefont {A.~J.}\ \bibnamefont {Sigillito}}, \bibinfo {author} {\bibfnamefont {M.~M.}\ \bibnamefont {Feldman}}, \bibinfo {author} {\bibfnamefont {E.}~\bibnamefont {Nielsen}},\ and\ \bibinfo {author} {\bibfnamefont {J.~R.}\ \bibnamefont {Petta}},\ }\href@noop {} {\bibfield  {journal} {\bibinfo  {journal} {Science Advances}\ }\textbf {\bibinfo {volume} {8}},\ \bibinfo {pages} {eabn5130} (\bibinfo {year} {2022})}\BibitemShut {NoStop}%
\bibitem [{\citenamefont {Sheldon}\ \emph {et~al.}(2016)\citenamefont {Sheldon}, \citenamefont {Magesan}, \citenamefont {Chow},\ and\ \citenamefont {Gambetta}}]{sheldon2016procedure}%
  \BibitemOpen
  \bibfield  {author} {\bibinfo {author} {\bibfnamefont {S.}~\bibnamefont {Sheldon}}, \bibinfo {author} {\bibfnamefont {E.}~\bibnamefont {Magesan}}, \bibinfo {author} {\bibfnamefont {J.~M.}\ \bibnamefont {Chow}},\ and\ \bibinfo {author} {\bibfnamefont {J.~M.}\ \bibnamefont {Gambetta}},\ }\href@noop {} {\bibfield  {journal} {\bibinfo  {journal} {Phys. Rev. A}\ }\textbf {\bibinfo {volume} {93}},\ \bibinfo {pages} {060302} (\bibinfo {year} {2016})}\BibitemShut {NoStop}%
\bibitem [{\citenamefont {Krantz}\ \emph {et~al.}(2019)\citenamefont {Krantz}, \citenamefont {Kjaergaard}, \citenamefont {Yan}, \citenamefont {Orlando}, \citenamefont {Gustavsson},\ and\ \citenamefont {Oliver}}]{krantz2019quantum}%
  \BibitemOpen
  \bibfield  {author} {\bibinfo {author} {\bibfnamefont {P.}~\bibnamefont {Krantz}}, \bibinfo {author} {\bibfnamefont {M.}~\bibnamefont {Kjaergaard}}, \bibinfo {author} {\bibfnamefont {F.}~\bibnamefont {Yan}}, \bibinfo {author} {\bibfnamefont {T.~P.}\ \bibnamefont {Orlando}}, \bibinfo {author} {\bibfnamefont {S.}~\bibnamefont {Gustavsson}},\ and\ \bibinfo {author} {\bibfnamefont {W.~D.}\ \bibnamefont {Oliver}},\ }\href@noop {} {\bibfield  {journal} {\bibinfo  {journal} {Applied physics reviews}\ }\textbf {\bibinfo {volume} {6}},\ \bibinfo {pages} {021318} (\bibinfo {year} {2019})}\BibitemShut {NoStop}%
\bibitem [{\citenamefont {Werninghaus}\ \emph {et~al.}(2021)\citenamefont {Werninghaus}, \citenamefont {Egger}, \citenamefont {Roy}, \citenamefont {Machnes}, \citenamefont {Wilhelm},\ and\ \citenamefont {Filipp}}]{werninghaus2021leakage}%
  \BibitemOpen
  \bibfield  {author} {\bibinfo {author} {\bibfnamefont {M.}~\bibnamefont {Werninghaus}}, \bibinfo {author} {\bibfnamefont {D.~J.}\ \bibnamefont {Egger}}, \bibinfo {author} {\bibfnamefont {F.}~\bibnamefont {Roy}}, \bibinfo {author} {\bibfnamefont {S.}~\bibnamefont {Machnes}}, \bibinfo {author} {\bibfnamefont {F.~K.}\ \bibnamefont {Wilhelm}},\ and\ \bibinfo {author} {\bibfnamefont {S.}~\bibnamefont {Filipp}},\ }\href@noop {} {\bibfield  {journal} {\bibinfo  {journal} {npj Quantum Information}\ }\textbf {\bibinfo {volume} {7}},\ \bibinfo {pages} {14} (\bibinfo {year} {2021})}\BibitemShut {NoStop}%
\bibitem [{\citenamefont {Egan}\ \emph {et~al.}(2021)\citenamefont {Egan}, \citenamefont {Debroy}, \citenamefont {Noel}, \citenamefont {Risinger}, \citenamefont {Zhu}, \citenamefont {Biswas}, \citenamefont {Newman}, \citenamefont {Li}, \citenamefont {Brown}, \citenamefont {Cetina} \emph {et~al.}}]{egan2021fault}%
  \BibitemOpen
  \bibfield  {author} {\bibinfo {author} {\bibfnamefont {L.}~\bibnamefont {Egan}}, \bibinfo {author} {\bibfnamefont {D.~M.}\ \bibnamefont {Debroy}}, \bibinfo {author} {\bibfnamefont {C.}~\bibnamefont {Noel}}, \bibinfo {author} {\bibfnamefont {A.}~\bibnamefont {Risinger}}, \bibinfo {author} {\bibfnamefont {D.}~\bibnamefont {Zhu}}, \bibinfo {author} {\bibfnamefont {D.}~\bibnamefont {Biswas}}, \bibinfo {author} {\bibfnamefont {M.}~\bibnamefont {Newman}}, \bibinfo {author} {\bibfnamefont {M.}~\bibnamefont {Li}}, \bibinfo {author} {\bibfnamefont {K.~R.}\ \bibnamefont {Brown}}, \bibinfo {author} {\bibfnamefont {M.}~\bibnamefont {Cetina}}, \emph {et~al.},\ }\href@noop {} {\bibfield  {journal} {\bibinfo  {journal} {Nature}\ }\textbf {\bibinfo {volume} {598}},\ \bibinfo {pages} {281} (\bibinfo {year} {2021})}\BibitemShut {NoStop}%
\bibitem [{\citenamefont {Pogorelov}\ \emph {et~al.}(2021)\citenamefont {Pogorelov}, \citenamefont {Feldker}, \citenamefont {Marciniak}, \citenamefont {Postler}, \citenamefont {Jacob}, \citenamefont {Krieglsteiner}, \citenamefont {Podlesnic}, \citenamefont {Meth}, \citenamefont {Negnevitsky}, \citenamefont {Stadler} \emph {et~al.}}]{pogorelov2021compact}%
  \BibitemOpen
  \bibfield  {author} {\bibinfo {author} {\bibfnamefont {I.}~\bibnamefont {Pogorelov}}, \bibinfo {author} {\bibfnamefont {T.}~\bibnamefont {Feldker}}, \bibinfo {author} {\bibfnamefont {C.~D.}\ \bibnamefont {Marciniak}}, \bibinfo {author} {\bibfnamefont {L.}~\bibnamefont {Postler}}, \bibinfo {author} {\bibfnamefont {G.}~\bibnamefont {Jacob}}, \bibinfo {author} {\bibfnamefont {O.}~\bibnamefont {Krieglsteiner}}, \bibinfo {author} {\bibfnamefont {V.}~\bibnamefont {Podlesnic}}, \bibinfo {author} {\bibfnamefont {M.}~\bibnamefont {Meth}}, \bibinfo {author} {\bibfnamefont {V.}~\bibnamefont {Negnevitsky}}, \bibinfo {author} {\bibfnamefont {M.}~\bibnamefont {Stadler}}, \emph {et~al.},\ }\href@noop {} {\bibfield  {journal} {\bibinfo  {journal} {PRX Quantum}\ }\textbf {\bibinfo {volume} {2}},\ \bibinfo {pages} {020343} (\bibinfo {year} {2021})}\BibitemShut {NoStop}%
\bibitem [{\citenamefont {Zeng}\ \emph {et~al.}(2018)\citenamefont {Zeng}, \citenamefont {Deng}, \citenamefont {Russo},\ and\ \citenamefont {Barnes}}]{zeng2018general}%
  \BibitemOpen
  \bibfield  {author} {\bibinfo {author} {\bibfnamefont {J.}~\bibnamefont {Zeng}}, \bibinfo {author} {\bibfnamefont {X.-H.}\ \bibnamefont {Deng}}, \bibinfo {author} {\bibfnamefont {A.}~\bibnamefont {Russo}},\ and\ \bibinfo {author} {\bibfnamefont {E.}~\bibnamefont {Barnes}},\ }\href@noop {} {\bibfield  {journal} {\bibinfo  {journal} {New J. Phys.}\ }\textbf {\bibinfo {volume} {20}},\ \bibinfo {pages} {033011} (\bibinfo {year} {2018})}\BibitemShut {NoStop}%
\bibitem [{\citenamefont {Zeng}\ and\ \citenamefont {Barnes}(2018)}]{zeng2018fastest}%
  \BibitemOpen
  \bibfield  {author} {\bibinfo {author} {\bibfnamefont {J.}~\bibnamefont {Zeng}}\ and\ \bibinfo {author} {\bibfnamefont {E.}~\bibnamefont {Barnes}},\ }\href@noop {} {\bibfield  {journal} {\bibinfo  {journal} {Phys. Rev. A}\ }\textbf {\bibinfo {volume} {98}},\ \bibinfo {pages} {012301} (\bibinfo {year} {2018})}\BibitemShut {NoStop}%
\bibitem [{\citenamefont {Zhuang}\ \emph {et~al.}(2022)\citenamefont {Zhuang}, \citenamefont {Zeng}, \citenamefont {Economou},\ and\ \citenamefont {Barnes}}]{zhuangNoiseresistantLandauZenerSweeps2022}%
  \BibitemOpen
  \bibfield  {author} {\bibinfo {author} {\bibfnamefont {F.}~\bibnamefont {Zhuang}}, \bibinfo {author} {\bibfnamefont {J.}~\bibnamefont {Zeng}}, \bibinfo {author} {\bibfnamefont {S.~E.}\ \bibnamefont {Economou}},\ and\ \bibinfo {author} {\bibfnamefont {E.}~\bibnamefont {Barnes}},\ }\href {https://doi.org/10.22331/q-2022-02-02-639} {\bibfield  {journal} {\bibinfo  {journal} {Quantum}\ }\textbf {\bibinfo {volume} {6}},\ \bibinfo {pages} {639} (\bibinfo {year} {2022})}\BibitemShut {NoStop}%
\bibitem [{\citenamefont {Dong}\ \emph {et~al.}(2021)\citenamefont {Dong}, \citenamefont {Zhuang}, \citenamefont {Economou},\ and\ \citenamefont {Barnes}}]{dongDoublyGeometricQuantum2021}%
  \BibitemOpen
  \bibfield  {author} {\bibinfo {author} {\bibfnamefont {W.}~\bibnamefont {Dong}}, \bibinfo {author} {\bibfnamefont {F.}~\bibnamefont {Zhuang}}, \bibinfo {author} {\bibfnamefont {S.~E.}\ \bibnamefont {Economou}},\ and\ \bibinfo {author} {\bibfnamefont {E.}~\bibnamefont {Barnes}},\ }\href {https://doi.org/10.1103/PRXQuantum.2.030333} {\bibfield  {journal} {\bibinfo  {journal} {PRX Quantum}\ }\textbf {\bibinfo {volume} {2}},\ \bibinfo {pages} {030333} (\bibinfo {year} {2021})}\BibitemShut {NoStop}%
\bibitem [{\citenamefont {Buterakos}\ \emph {et~al.}(2021)\citenamefont {Buterakos}, \citenamefont {Das~Sarma},\ and\ \citenamefont {Barnes}}]{buterakosGeometricalFormalismDynamically2021}%
  \BibitemOpen
  \bibfield  {author} {\bibinfo {author} {\bibfnamefont {D.}~\bibnamefont {Buterakos}}, \bibinfo {author} {\bibfnamefont {S.}~\bibnamefont {Das~Sarma}},\ and\ \bibinfo {author} {\bibfnamefont {E.}~\bibnamefont {Barnes}},\ }\href {https://doi.org/10.1103/PRXQuantum.2.010341} {\bibfield  {journal} {\bibinfo  {journal} {PRX Quantum}\ }\textbf {\bibinfo {volume} {2}},\ \bibinfo {pages} {010341} (\bibinfo {year} {2021})}\BibitemShut {NoStop}%
\bibitem [{\citenamefont {Barnes}\ \emph {et~al.}(2022)\citenamefont {Barnes}, \citenamefont {{Calderon-Vargas}}, \citenamefont {Dong}, \citenamefont {Li}, \citenamefont {Zeng},\ and\ \citenamefont {Zhuang}}]{barnesDynamicallyCorrectedGates2022}%
  \BibitemOpen
  \bibfield  {author} {\bibinfo {author} {\bibfnamefont {E.}~\bibnamefont {Barnes}}, \bibinfo {author} {\bibfnamefont {F.~A.}\ \bibnamefont {{Calderon-Vargas}}}, \bibinfo {author} {\bibfnamefont {W.}~\bibnamefont {Dong}}, \bibinfo {author} {\bibfnamefont {B.}~\bibnamefont {Li}}, \bibinfo {author} {\bibfnamefont {J.}~\bibnamefont {Zeng}},\ and\ \bibinfo {author} {\bibfnamefont {F.}~\bibnamefont {Zhuang}},\ }\href {https://doi.org/10.1088/2058-9565/ac4421} {\bibfield  {journal} {\bibinfo  {journal} {Quantum Sci. Technol.}\ }\textbf {\bibinfo {volume} {7}},\ \bibinfo {pages} {023001} (\bibinfo {year} {2022})}\BibitemShut {NoStop}%
\bibitem [{\citenamefont {Hai}\ \emph {et~al.}(2022)\citenamefont {Hai}, \citenamefont {Li}, \citenamefont {Zeng},\ and\ \citenamefont {Deng}}]{hai2022universal}%
  \BibitemOpen
  \bibfield  {author} {\bibinfo {author} {\bibfnamefont {Y.-J.}\ \bibnamefont {Hai}}, \bibinfo {author} {\bibfnamefont {J.}~\bibnamefont {Li}}, \bibinfo {author} {\bibfnamefont {J.}~\bibnamefont {Zeng}},\ and\ \bibinfo {author} {\bibfnamefont {X.-H.}\ \bibnamefont {Deng}},\ }\href@noop {} {\bibfield  {journal} {\bibinfo  {journal} {arXiv:2210.14521}\ } (\bibinfo {year} {2022})}\BibitemShut {NoStop}%
\bibitem [{\citenamefont {Wallman}\ and\ \citenamefont {Emerson}(2016)}]{wallmanNoiseTailoringScalable2016}%
  \BibitemOpen
  \bibfield  {author} {\bibinfo {author} {\bibfnamefont {J.~J.}\ \bibnamefont {Wallman}}\ and\ \bibinfo {author} {\bibfnamefont {J.}~\bibnamefont {Emerson}},\ }\href {https://doi.org/10.1103/PhysRevA.94.052325} {\bibfield  {journal} {\bibinfo  {journal} {Phys. Rev. A}\ }\textbf {\bibinfo {volume} {94}},\ \bibinfo {pages} {052325} (\bibinfo {year} {2016})}\BibitemShut {NoStop}%
\bibitem [{\citenamefont {Hashim}\ \emph {et~al.}(2021)\citenamefont {Hashim}, \citenamefont {Naik}, \citenamefont {Morvan}, \citenamefont {Ville}, \citenamefont {Mitchell}, \citenamefont {Kreikebaum}, \citenamefont {Davis}, \citenamefont {Smith}, \citenamefont {Iancu}, \citenamefont {O'Brien}, \citenamefont {Hincks}, \citenamefont {Wallman}, \citenamefont {Emerson},\ and\ \citenamefont {Siddiqi}}]{hashimRandomizedCompilingScalable2021}%
  \BibitemOpen
  \bibfield  {author} {\bibinfo {author} {\bibfnamefont {A.}~\bibnamefont {Hashim}}, \bibinfo {author} {\bibfnamefont {R.~K.}\ \bibnamefont {Naik}}, \bibinfo {author} {\bibfnamefont {A.}~\bibnamefont {Morvan}}, \bibinfo {author} {\bibfnamefont {J.-L.}\ \bibnamefont {Ville}}, \bibinfo {author} {\bibfnamefont {B.}~\bibnamefont {Mitchell}}, \bibinfo {author} {\bibfnamefont {J.~M.}\ \bibnamefont {Kreikebaum}}, \bibinfo {author} {\bibfnamefont {M.}~\bibnamefont {Davis}}, \bibinfo {author} {\bibfnamefont {E.}~\bibnamefont {Smith}}, \bibinfo {author} {\bibfnamefont {C.}~\bibnamefont {Iancu}}, \bibinfo {author} {\bibfnamefont {K.~P.}\ \bibnamefont {O'Brien}}, \bibinfo {author} {\bibfnamefont {I.}~\bibnamefont {Hincks}}, \bibinfo {author} {\bibfnamefont {J.~J.}\ \bibnamefont {Wallman}}, \bibinfo {author} {\bibfnamefont {J.}~\bibnamefont {Emerson}},\ and\ \bibinfo {author} {\bibfnamefont {I.}~\bibnamefont {Siddiqi}},\ }\href {https://doi.org/10.1103/PhysRevX.11.041039} {\bibfield  {journal} {\bibinfo  {journal} {Phys.
  Rev. X}\ }\textbf {\bibinfo {volume} {11}},\ \bibinfo {pages} {041039} (\bibinfo {year} {2021})}\BibitemShut {NoStop}%
\bibitem [{\citenamefont {Gu}\ \emph {et~al.}(2022)\citenamefont {Gu}, \citenamefont {Ma}, \citenamefont {Forcellini},\ and\ \citenamefont {Liu}}]{guNoiseresilientPhaseEstimation2022}%
  \BibitemOpen
  \bibfield  {author} {\bibinfo {author} {\bibfnamefont {Y.}~\bibnamefont {Gu}}, \bibinfo {author} {\bibfnamefont {Y.}~\bibnamefont {Ma}}, \bibinfo {author} {\bibfnamefont {N.}~\bibnamefont {Forcellini}},\ and\ \bibinfo {author} {\bibfnamefont {D.~E.}\ \bibnamefont {Liu}},\ }\href@noop {} {\bibfield  {journal} {\bibinfo  {journal} {arXiv:2208.04100}\ } (\bibinfo {year} {2022})},\ \Eprint {https://arxiv.org/abs/2208.04100} {arxiv:2208.04100} \BibitemShut {NoStop}%
\bibitem [{\citenamefont {Urbanek}\ \emph {et~al.}(2021)\citenamefont {Urbanek}, \citenamefont {Nachman}, \citenamefont {Pascuzzi}, \citenamefont {He}, \citenamefont {Bauer},\ and\ \citenamefont {de~Jong}}]{urbanek2021mitigating}%
  \BibitemOpen
  \bibfield  {author} {\bibinfo {author} {\bibfnamefont {M.}~\bibnamefont {Urbanek}}, \bibinfo {author} {\bibfnamefont {B.}~\bibnamefont {Nachman}}, \bibinfo {author} {\bibfnamefont {V.~R.}\ \bibnamefont {Pascuzzi}}, \bibinfo {author} {\bibfnamefont {A.}~\bibnamefont {He}}, \bibinfo {author} {\bibfnamefont {C.~W.}\ \bibnamefont {Bauer}},\ and\ \bibinfo {author} {\bibfnamefont {W.~A.}\ \bibnamefont {de~Jong}},\ }\href@noop {} {\bibfield  {journal} {\bibinfo  {journal} {Phys. Rev. Lett.}\ }\textbf {\bibinfo {volume} {127}},\ \bibinfo {pages} {270502} (\bibinfo {year} {2021})}\BibitemShut {NoStop}%
\bibitem [{\citenamefont {Cai}\ and\ \citenamefont {Benjamin}(2019)}]{cai2019constructing}%
  \BibitemOpen
  \bibfield  {author} {\bibinfo {author} {\bibfnamefont {Z.}~\bibnamefont {Cai}}\ and\ \bibinfo {author} {\bibfnamefont {S.~C.}\ \bibnamefont {Benjamin}},\ }\href@noop {} {\bibfield  {journal} {\bibinfo  {journal} {Scientific reports}\ }\textbf {\bibinfo {volume} {9}},\ \bibinfo {pages} {1} (\bibinfo {year} {2019})}\BibitemShut {NoStop}%
\end{thebibliography}%

\end{document}

\preprint{APS/123-QED}

\title{Supplementary Materials: Quantum Circuit noise robustness from a geometric perspective}% Force line breaks with \\
%\thanks{A footnote to the article title}%
\author{Junkai Zeng}
\email{zengjk@sustech.edu.cn}
\affiliation{Shenzhen Institute for Quantum Science and Engineering (SIQSE), Southern University of Science and Technology, Shenzhen, P. R. China}%
\affiliation{International Quantum Academy (SIQA), and Shenzhen Branch, Hefei National Laboratory, Futian District, Shenzhen, P. R. China}
\author{Yong-Ju Hai}
\affiliation{International Quantum Academy (SIQA), and Shenzhen Branch, Hefei National Laboratory, Futian District, Shenzhen, P. R. China}
\author{Hao Liang}
\affiliation{Shenzhen Institute for Quantum Science and Engineering (SIQSE), Southern University of Science and Technology, Shenzhen, P. R. China}%
\affiliation{International Quantum Academy (SIQA), and Shenzhen Branch, Hefei National Laboratory, Futian District, Shenzhen, P. R. China}
\author{Xiu-Hao Deng}
\email{dengxh@sustech.edu.cn}
\affiliation{Shenzhen Institute for Quantum Science and Engineering (SIQSE), Southern University of Science and Technology, Shenzhen, P. R. China}%
\affiliation{International Quantum Academy (SIQA), and Shenzhen Branch, Hefei National Laboratory, Futian District, Shenzhen, P. R. China}

\maketitle

\section{Distribution of Randomly Twirled Error Operator}
Here we derive the randomly twirled error path distribution. We consider only Pauli twirling, as more complex twirling will typically require even more complex correction operators that are hard to implement. The total twirled error operator can be expressed as:
\begin{equation}
    \begin{split}
        \tilde{\Phi}_{\text{total}} &= \sum_{i=0}^{N_D} \sum_j \gamma^{[i]}_j \Gamma_{\text{lo},j}^{[i]}\left(\mathcal{C}^{[i]\dagger}\sigma_j\mathcal{C}^{[i]}\right)\\ 
&= \sum_{i=0}^{N_D} \sum_{j,k} \gamma^{[i]}_j \Gamma_{\text{lo},j}^{[i]}\mathcal{R}^{[i],j}_k \sigma_k
    \end{split}
\end{equation}
where $\gamma^{[i]}_j=1$ if the twirling operator at layer $i$ and $\sigma_j$ commutes, and -1 if they anti-commutes. $\mathcal{R}^{[i]}$ is the Choi matrix representation for the unitary transformation $\mathcal{C}^{[i]}$ that satisfy $|\mathcal{R}^{[i],j}|=1$ for arbitrary $j$. Since each of the $\gamma_j^{[i]}$ terms is independent, the $k$-component of $\tilde{\Phi}_{\text{total}}$ converges towards a zero-mean Gaussian distribution as per the Lindeberg central limit theorem. The variance is given by $\sum_i^{N_D}\sum_j \left(\Gamma_{\text{lo},j}^{[i]}\mathcal{R}^{[i],j}_k\right)^2$. Particularly, the expected squared error distance from the origin is calculated as

\begin{equation}
\begin{split}
    \langle |\tilde{\Phi}_{\text{total}}|^2\rangle &= \sum_k \sum_{i=0}^{N_D}\sum_j \left(\Gamma_{\text{lo},j}^{[i]}\mathcal{R}^{[i],j}_k \right)^2 \\
    & = \sum_{i=0}^{N_D}\sum_j \left(\Gamma_{\text{lo},j}^{[i]}\right)^2\\
    &=\sum_{i=0}^{N_D} \left|\Gamma^{[i]}\right|^2=N_D \Gamma^2
\end{split}
\end{equation}
This shows the statement of $\langle|\Phi_{\text{total}}|\rangle \sim \sqrt{N_{D}}\Gamma$ in the main body.

\section{Computing Error Coefficients $\Gamma^{n}$}
In the main part of the manuscript, we did not consider the inner structure of the error model for simplification. 
Here we make a brief discussion on obtaining detailed information on the parameters of the noise channel. 
For a general time-dependent noise, the first order error in the Magnus expansion of the error operator is $\Gamma^{1}_j=\sum_i R_j^{\epsilon(i)}(T)=\sum_i \int_0^T \epsilon_i(t) r_{i,j}'(t)dt$. Here we use $\Gamma_j^n$ to represent $n$'th order Magnus expansion for a circuit layer, and use $R_j^{\epsilon(i)}(t)$ for the corresponding continuous dynamics of the gate implementation. 
When the error coefficient $\Gamma_j^1$ appears in the noise channel after performing randomized compiling, it only appears in the second order term as its variance $\langle \left(\Gamma_j^{1}\right)^2\rangle$, which is given by 
\begin{equation}
\label{eq:channelavg}
\begin{split}
    \langle R_{j}^{\epsilon(i)}(t) \rangle = \int_0^t \bar{\epsilon}_i(s) r'_{i,j}(s)ds \\ 
    \langle \left( \Gamma_j^{1} \right)^2\rangle =  \sum_{ik}\langle R_{j}^{\epsilon(i)}(T) \rangle  \langle R_{j}^{\epsilon(k)}(T) \rangle\\ 
    + \sum_{ik}\int_{-\infty}^{\infty} S_{ik}(\omega) F_{ij}^{(1)*}(\omega,T)F_{kj}^{(1)}(\omega,T) d\omega\\ 
\end{split}
\end{equation}
here the 1st-order \emph{fundamental filter function} is defined as $F_{ij}^{(1)}(\omega,t)=\int_0^t C_\epsilon^{(i)}(s) r_{i,j}'(s)e^{i \omega s} ds$ and $C_\epsilon^{(i)}$ is a time-dependent factor that accounts for changes in noise amplitude over time. 
For example, in the case of multiplicative amplitude noise, we have $C_\epsilon(t)=\Omega(t)$, and $\epsilon_\Omega(t)=\tilde{\epsilon}(t) \Omega(t)$ where $\tilde{\epsilon}(t)$ is a stationary stochastic process. 
The power spectral density (PSD) $S_{ik}(\omega)=\frac{1}{2\pi}\int_{-\infty}^{\infty} ds e^{-i \omega s} \langle \tilde{\epsilon}_i(s) \tilde{\epsilon}_k(0)\rangle$. 
While this is a general formula, in many cases the noise can be assumed to be stationary with zero mean (where $\langle R_{j}^{\epsilon(i)}(t) \rangle$ becomes zero) or quasistatic where $\Gamma^{1}_j=\sum_i \epsilon_i r_{i,j}(T)dt$, and a geometric curve (QEED) can fully describe the dynamic of the qubit. 

Although the same approach can be applied to higher-order terms, their expressions can become more complex as high-order correlations are involved, and we will not discuss them here.

\section{Perturbative Expansion Of The Noise Channel}
\label{app:noisech}
In this section we derive the analytical formula for the noise channel when RC is applied to the circuit.
We have in general
\begin{equation}
\begin{split}
    \mathcal{E}(P)&=\exp{-i \Phi} P \exp{i \Phi}\\
    &\Phi=\sum_{n=1}^{\infty} \frac{1}{n!}\sum_i \epsilon^n \Gamma_i^{n} \sigma_i
    %\mathcal{E}(\hat{P})&=\exp\left(-i \sum_{n=1}^{\infty} \frac{1}{n!}\sum_i \epsilon^n R_i^{[n]} \hat{\sigma}_i\right)\hat{P} \exp\left(i \sum_{n=1}^{\infty} \frac{1}{n!}\sum_i \epsilon^n R_i^{[n]} \hat{\sigma}_i\right)\\
    %&=\hat{P}\exp\left(-i \sum_{n=1}^{\infty} \frac{1}{n!}\sum_i \epsilon^n R_i^{[n]} \hat{P}\hat{\sigma}_i\hat{P}\right) \exp\left(i \sum_{n=1}^{\infty} \frac{1}{n!}\sum_i \epsilon^n R_i^{[n]} \hat{\sigma}_i\right)
\end{split}
\end{equation}
where $\Gamma_i^{n}$ is the n-th order error coefficient calculated through Magnus expansion. We can expand the expression using the Baker-Campbell-Hausdorff formula:
\begin{equation}
    \begin{split}
        \mathcal{E}(P)&=\sum_{n=0}^{\infty}\mathcal{E}_{\text{BCH}}^{n}(P)\\
        \mathcal{E}_{\text{BCH}}^{n}(P)&=\frac{(-i)^n}{n!}%\sum_{i_1,i_2,...,i_n}\sum_{m_1,m_2,...,m_n}
        \sum_{\substack{i_1,i_2,...,i_n\\m_1,m_2,...,m_n}
                                    }\prod_{j=1}^n\frac{\Gamma_{i_j}^{m_j}}{m_j!}[\sigma_{i_1}, [\sigma_{i_2},...[\sigma_{i_n},P]]...]
    \end{split}
\end{equation}
However, we are more interested in the expansion about the error strength $\epsilon$: $\mathcal{E}(P) = \sum_n^{\infty} \mathcal{E}_\epsilon^{n}(P)$ with each term $\mathcal{E}_\epsilon^{n}(P)$ has the same order of magnitude as $\epsilon^n$. Further, write the expansion to get: 
\begin{equation}
    \mathcal{E}^{n}_\epsilon(P) = \mathcal{E}^{n,1}_\epsilon(P) + \mathcal{E}^{n,2}_\epsilon(P)+...+\mathcal{E}^{n,n}_\epsilon(P)  
\end{equation}
whereas each $\mathcal{E}^{n,k}(P)$ is the contribution of n-th order term of the expanded noise channel contributed by BCH expansion $\mathcal{E}_{\text{BCH}}^{k}(P)$.

This can be further expressed as 
\begin{equation}
    \begin{split}
        &\mathcal{E}^{n,k}(P)=\frac{(-i)^k}{k!}\sum_{i_1,i_2,...,i_k}\mathcal{K}_{i_1,i_2,...i_k}^{n}\mathcal{P}_{i_1,i_2,...i_k}\\
        &\mathcal{K}_{i_1,i_2,...i_k}^{n} = \sum_{\sum_j m_j=n}\frac{\Gamma_{i_1}^{m_1} \Gamma_{i_2}^{m_2}... \Gamma_{i_k}^{m_k}}{m_1! m_2!...m_k!}\\
        &\mathcal{P}_{i_1,i_2,...i_k} = [\sigma_{i_2},...[\sigma_{i_n},P]]...]
    \end{split}
\end{equation}

It's easy to see that $\mathcal{K}$ has a symmetrical feature on all its indices: $\mathcal{K}_{...i_\alpha,...,i_\beta,...}^{n}=\mathcal{K}_{...i_\beta,...,i_\alpha,...}^{n}$. Meanwhile, for $\mathcal{P}$, reordering $i$'s will add a plus or minus sign before it, making $\mathcal{P}$ either symmetrical or anti-symmetrical about its indices.
As a result, the summation of $\mathcal{K}_{i_1,i_2,...i_k}^{n}\mathcal{P}_{i_1,i_2,...i_k}$ can be reduced to cases where $\mathcal{P}$ exhibits the same symmetry, as all combinations of $\sigma$'s that making $\mathcal{P}$ anti-symmetrical will vanish. 
This requires $[\sigma_{i_\alpha}, \sigma_{i_\beta}]=0$ for all pairs of indices $(i_\alpha, i_\beta)$ in $\mathcal{P}$. This also requires that all $\sigma_i$ in $\mathcal{P}$ anti-commute with $P$ as otherwise $\mathcal{P}$ becomes zero. In this case, $\mathcal{P}_{i_1,i_2,...i_k}=2^k \sigma_{i_1}\sigma_{i_2}...\sigma_{i_k}P$.

Now consider cases when RC is applied. With random twirling gate $T$, all Pauli operators appear in $\mathcal{P}$ becomes $T^\dagger \sigma T$ and therefore we have $\tilde{\mathcal{P}}_{i_1,i_2,...i_k} = 2^k T^\dagger \sigma_{i_1}\sigma_{i_2}...\sigma_{i_n} T P$. 
Thus, after averaging multiple shots over randomly chosen twirling gates, we obtain $\mathbb{E}_T(T^\dagger \sigma_{i_1}\sigma_{i_2}...\sigma_{i_n} T)=0$ for $\sigma_{i_1}\sigma_{i_2}...\sigma_{i_n} \neq \mathcal{I}$. 
Consequently, for most of the index configurations of $\mathcal{P}$, we get a zero after performing the average, except for cases where the Pauli product chain forms an identity operator, giving another constraint on the combination of indices of $\mathcal{K}^{n}$ and $\mathcal{P}$. 
The resulting noise channel, after randomized compiling, is then transformed from a coherent noise channel to a Pauli channel, as we have $\mathcal{P}\propto P$. 

The constraints on the Pauli product chain also affect the number of $k$. 
Note that we always have $[\sigma_{i_1}\sigma_{i_2}...\sigma_{i_{k-1}}, P]=0$ for odd $k$, and $\sigma_{i_{k}}=\pm\sigma_{i_1}\sigma_{i_2}...\sigma_{i_{k-1}}$ must be satisfied for the entire product chain to equal the identity. 
This conflicts with the requirement that $\sigma_{i_{k}}$ shall anticommute with $P$, and therefore $k$ must be even.

\bibliography{apssamp}% Produces the bibliography via BibTeX.